\documentclass[a4paper,11pt]{article}
\usepackage{jheppub}
\usepackage{amsmath,amsfonts,amssymb}
\usepackage{xcolor}
\usepackage{graphicx,float}
\usepackage[utf8]{inputenc}
\usepackage{hyperref}
\title{Entanglement islands and information recovery from near-extremal regular black holes}
\author{Ankit Anand$^1$,}
\emailAdd{anand@iitk.ac.in}
\affiliation{$^1$Department of Physics, Indian Institute of Technology, Kanpur 208016, India.}
\author{Kimet Jusufi$^2$,}
\emailAdd{kimet.jusufi@unite.edu.mk}
\affiliation{$^2$Physics Department, State University of Tetovo, Ilinden Street nn, 1200, Tetovo, North Macedonia.}
\author{Amir A.~Khodahami$^{3,4}$ and}
\emailAdd{a.khodahami@shirazu.ac.ir}
\author{Ahmad Sheykhi$^{3,4}$}
\emailAdd{asheykhi@shirazu.ac.ir}
\affiliation{$^{3}$Department of Physics, College of Science, Shiraz
	University, Shiraz 71454, Iran \\ $^{4}$Biruni Observatory, College of
	Science, Shiraz University, Shiraz 71454, Iran}
\abstract{
	We investigate the Page curve and information recovery in a near-extremal regular black hole inspired by T-duality, in which the central singularity is resolved by a minimal length scale. By integrating the first law of thermodynamics at fixed minimal length, we obtain a black-hole entropy containing an intrinsically quantum logarithmic correction, while the area-law contribution vanishes in the extremal limit. Consequently, the extremal remnant carries a finite entropy of purely quantum origin. Using the island prescription in the near-horizon regime, we evaluate the generalized entropy of radiation and compare the standard area functional with an alternative functional constructed from the corrected thermodynamic entropy of the black hole. In the near-extremal limit, the latter reduces analytically to the classical extremization problem with an effective coupling, and it places the physical island closer to the outer horizon. Although both prescriptions produce a Page transition, they predict different saturation values and Page times. While the standard functional yields a plateau controlled by the area term, the corrected prescription saturates at the full thermodynamic entropy of the outer horizon and, in the extremal limit, at the logarithmic remnant entropy. Including evaporation and backreaction, the Page curve develops the expected descending branch and asymptotes to the entropy of the cold extremal remnant rather than to zero. Our results indicate that a thermodynamically consistent description of information recovery from regular near-extremal black holes requires incorporating the intrinsic quantum correction to the gravitational entropy.
}
\begin{document}
%%%%%%%%%%%%%%%%%%%%%%%%%%%%%%%%%%%%%%%%%%%%%%%%%%%%%%%%%%%%%
\maketitle
%%%%%%%%%%%%%%%%%%%%%%%%%%%%%%%%%%%%%%%%%%%%%%%%%%%%%%%%%%%%%
\section{Introduction}
\label{sec:intro}
The black hole information paradox~\cite{Hawking:1976ra} remains one of the sharpest
open problems in quantum gravity. Hawking's semiclassical
calculation~\cite{Hawking:1975vcx} implies that the entanglement entropy
of the radiation emitted by an eternal black hole grows monotonically without bound, in
manifest tension with the unitary evolution demanded by quantum mechanics. Unitarity
instead requires that this entropy follow the Page curve~\cite{Page:1993df,Page:2013dx},
rising until roughly half the coarse-grained entropy has been radiated and decreasing
thereafter.

Over the past few years this behaviour has been reproduced within semiclassical gravity
by the island prescription~\cite{Penington:2019npb,Almheiri:2019psf,Almheiri:2019hni, Almheiri:2019yqk}, whose validity is supported by replica wormhole
computations~\cite{Almheiri:2020cfm}. The underlying quantum extremal surface
prescription originates in holography~\cite{Ryu:2006bv,Hubeny:2007xt}, and the island
rule was established in two-dimensional gravity using the replica
trick~\cite{Holzhey:1994we,Callan:1994py,Calabrese:2009qy} and interpreted in terms of
replica wormholes~\cite{Penington:2019kki,Almheiri:2019qdq}. Its realisation in JT
gravity and related low-dimensional models has produced an extensive literature on
islands and Page curves~\cite{Hollowood:2020cou,Ageev:2019xii}, including asymptotically
flat two-dimensional geometries~\cite{Gautason:2020tmk,Anegawa:2020ezn,Hartman:2020swn,
RoyChowdhury:2022awr}, and the prescription has since been extended to asymptotically
flat black holes in four and higher dimensions and to other
backgrounds~\cite{Chen:2019uhq,Hashimoto:2020cas,Wang:2021woy,Matsuo:2020ypv,
Raju:2020smc,Alishahiha:2020qza,Anand:2022mla,Anand:2025gsb,Anand:2023ozw,Yadav:2022jib,
Yadav:2022fmo,Goswami:2023ovb,Uhlemann:2021nhu,He:2021mst,Wang:2021mqq}, with related
developments in stringy axion systems and beyond-$AdS_2$
settings~\cite{Choudhury:2022mch,Choudhury:2017qyl,Choudhury:2017bou,Krishnan:2020oun}.
The great majority of these analyses concern
singular black holes, for which the endpoint of evaporation lies outside the regime of
validity of the calculation. Regular black holes, in which the central singularity is
resolved by a minimal length scale of quantum-gravitational origin, therefore provide a
natural laboratory in which the late stages of evaporation and the ultimate fate of the
information can be followed further. The Bardeen solution~\cite{Bardeen1968} and its
generalisations~\cite{Hayward2006} are the classic examples; here we work with the
 regular black hole inspired by T-duality of Ref.~\cite{nicolini2019}, characterized by a
minimal length $\ell_0$. Regular black hole models with T-duality have been investigated also in  \cite{Gaete:2022ukm,Jusufi:2024dtr,Jusufi:2025qgd}. 

Two features make this model particularly instructive. First, integrating the first law
at fixed $\ell_0$ yields an entropy that is \emph{not} one quarter of the horizon area
but carries an additional logarithmic term proportional to $\ell_0^{2}$. Logarithmic
corrections of this kind are ubiquitous in quantum
gravity~\cite{Solodukhin:2011gn,karan2021,sen2012,sen2012a,banerjee2011,banerjee2011a,
banerjee2021,iliesiu2025a} and are known to dominate for extremal
black holes. Second, in the present model the area-law piece vanishes identically at
extremality, so the entire entropy of the extremal configuration is logarithmic. The
extremal remnant is thus a state whose entropy is purely quantum-gravitational in origin.

This dual structure raises a question that, to our knowledge, has not been addressed
before. The island prescription instructs one to extremize the \emph{gravitational
entropy} associated with the entangling surface; in the standard treatment this is taken
to be the area term $\mathrm{Area}(\partial I)/4G$. When the theory itself assigns a
different entropy to a sphere of given radius --- as it does here --- it is natural to ask
whether the island functional should be built from that corrected entropy instead. In this
paper we carry out both computations and compare them. We find that the corrected
functional is analytically tractable: the variation of the corrected entropy is an
algebraic function of the island radius, which reduces the extremization problem to the
classical one with a single rescaled coupling, and which fixes the extremal Page plateau
to be exactly the logarithmic entropy $S_{\rm ext}$ rather than the extremal horizon area.
The two prescriptions therefore differ in a way that is directly tied to the physics of the
remnant.

The paper is organised as follows. Section~\ref{sec:setup} fixes our conventions for the
island prescription and for the matter entanglement entropy after spherical reduction.
Section~\ref{sec:geometry} reviews the geometry and thermodynamics of the T-duality black
hole, derives the corrected entropy and shows that the near-horizon geometry of the
extremal solution is $AdS_2\times S^2$. Section~\ref{sec:no_island} computes the
no-island radiation entropy in Kruskal-like coordinates.
Section~\ref{sec:island} performs the island computation with the standard area
functional, and Section~\ref{sec:island_quantum} repeats it with the corrected entropy
functional. Section~\ref{sec:page} assembles the Page curves for both prescriptions.
Section~\ref{sec:discussion} discusses the implications for the information paradox and
Section~\ref{sec:conclusion} concludes.

%%%%%%%%%%%%%%%%%%%%%%%%%%%%%%%%%%%%%%%%%%%%%%%%%%%%%%%%%%%%%
\section{Set-up and conventions}
\label{sec:setup}

The entropy of a region $R$ of Hawking radiation is computed by extremizing the
generalized entropy over candidate island regions $I$,
\begin{equation}
	\label{eq:island}
	S(R) = \min_{\rm ext} \left[
	\frac{\mathrm{Area}(\partial I)}{4G_N}
	+ S_{\mathrm{matter}}(R \cup I)
	\right].
\end{equation}
In higher dimensions, the matter entanglement entropy contains UV divergences localized
near the entangling surfaces. Schematically,
\begin{equation}
	S_{\mathrm{matter}}(R \cup I)
	=
	\frac{\mathrm{Area}(\partial I)}{\epsilon_{\rm uv}^{2}}
	+
	S_{\mathrm{matter}}^{\mathrm{(finite)}}(R \cup I),
\end{equation}
where $\epsilon_{\rm uv}$ is a short-distance cutoff. As in
\cite{Almheiri:2020cfm}, this area divergence can be absorbed into a renormalization of
Newton's constant,
\begin{equation}
	\frac{1}{4G(\partial I)}
	=
	\frac{1}{4G_N}
	+
	\frac{1}{\epsilon_{\rm uv}^{2}},
\end{equation}
leading to the renormalized prescription
\begin{equation}\label{eq:renormalized_prescription}
	S(R) = \min_{\rm ext} \left[
	\frac{\mathrm{Area}(\partial I)}{4G(\partial I)}
	+
	S_{\mathrm{matter}}^{\mathrm{(finite)}}(R \cup I)
	\right].
\end{equation}

For an eternal black hole without an island, $I=\emptyset$, the gravitational area term is
absent and the radiation region consists of two exterior components,
$R=R_+\cup R_-$. The time dependence of the entropy is controlled by the correlations
between the two components, equivalently by the mutual information $I(R_+;R_-)$.
Up to time-independent local terms, the finite part of the entropy may be written as
\begin{equation}\label{eq:no_island_mutual_info}
	S_{\mathrm{matter}}^{\mathrm{(finite)}}(R)
	=
	\text{const.}
	-
	I(R_+;R_-).
\end{equation}
As the two radiation regions become more widely separated, their mutual information
decreases, and the no-island entropy correspondingly increases.

In higher-dimensional geometries the matter sector obtained after spherical reduction
contains a Kaluza--Klein tower of massive modes, so the mutual information between two
spacelike-separated regions is sensitive to the separation scale. Following
Ref.~\cite{Hashimoto:2020cas} we use two limiting regimes.

When the separation between the boundaries is much larger than the correlation lengths of
the massive KK modes, only the s-wave sector remains effectively massless and the system
reduces to a two-dimensional CFT of central charge $c$, for which
\begin{equation}\label{eq:MI_large_sep}
I(A;B)=-\frac{c}{3}\log\left(\frac{d(x,y)}{\epsilon_{\rm uv}}\right),
\end{equation}
where $x$ and $y$ are the boundaries of $A$ and $B$ and $d(x,y)$ is the conformal distance
between them.

When instead the two boundary surfaces are parallel and separated by a small proper
distance $L$, the mutual information is dominated by short-distance correlations and takes
the universal form
\begin{equation}
I(A;B)=\chi\,c\,\frac{\mathrm{Area}}{L^{2}},
\label{eq:Information-Area}
\end{equation}
for $c$ free massless fields, with $\chi$ a pure number. In four spacetime dimensions the
numerical evaluation quoted in Ref.~\cite{Hashimoto:2020cas} gives $\chi=0.00554$ for a boson and
$\chi=0.00538$ for a fermion.%
%%% TODO: add the primary numerical reference for chi (values quoted via Hashimoto:2020cas).

Although \eqref{eq:Information-Area} is derived in flat
space, it applies locally whenever the curvature radius is large compared with $L$.

The first regime will be used in the no-island analysis, where the two radiation endpoints
are separated by a large conformal distance; the second will be used in the island phase,
where the island boundary lies close to the radiation boundary. Since the regular black
hole considered here is spherically symmetric and admits a consistent reduction to an
effective two-dimensional theory in the near-horizon region, both approximations may be
taken over unchanged.

It is convenient to introduce once and for all the dimensionless combination
\begin{equation}
\lambda\;\equiv\;\frac{4\chi c\,G}{\bar{\ell}_0^{2}},
\label{eq:lambda_def}
\end{equation}
where $\bar{\ell}_0$ is the $AdS_2$ radius defined in Sec.~\ref{sec:geometry}. This is the
only combination in which the matter central charge enters the island analysis;
$\lambda\ll1$ is the statement that the near-horizon curvature radius is large in Planck
units, which is precisely the regime in which the semiclassical treatment is justified. All
numerical illustrations below use the representative value $\lambda=10^{-2}$.

Unless stated otherwise we work in units $\hbar=k_B=G=1$ and measure all lengths in units
of the extremal radius, $r_e=1$, keeping Newton's constant explicit: it is the ratio
$G/r_e^{2}$ that controls the semiclassical expansion, and setting it to unity would place
the horizon at the Planck scale. The numerical illustrations below use $\lambda=10^{-2}$
with $c=100$, which by Eq.~\eqref{eq:lambda_def} corresponds to
$G/r_e^{2}\simeq6.8\times10^{-3}$, i.e.\ an extremal horizon some twelve Planck lengths
across --- comfortably semiclassical, yet small enough that the logarithmic correction is
not negligible. Throughout, $c$ denotes the central charge of the effective
two-dimensional matter sector and never the speed of light.
\begin{figure}[t]
	\centering
	\includegraphics[width=\textwidth]{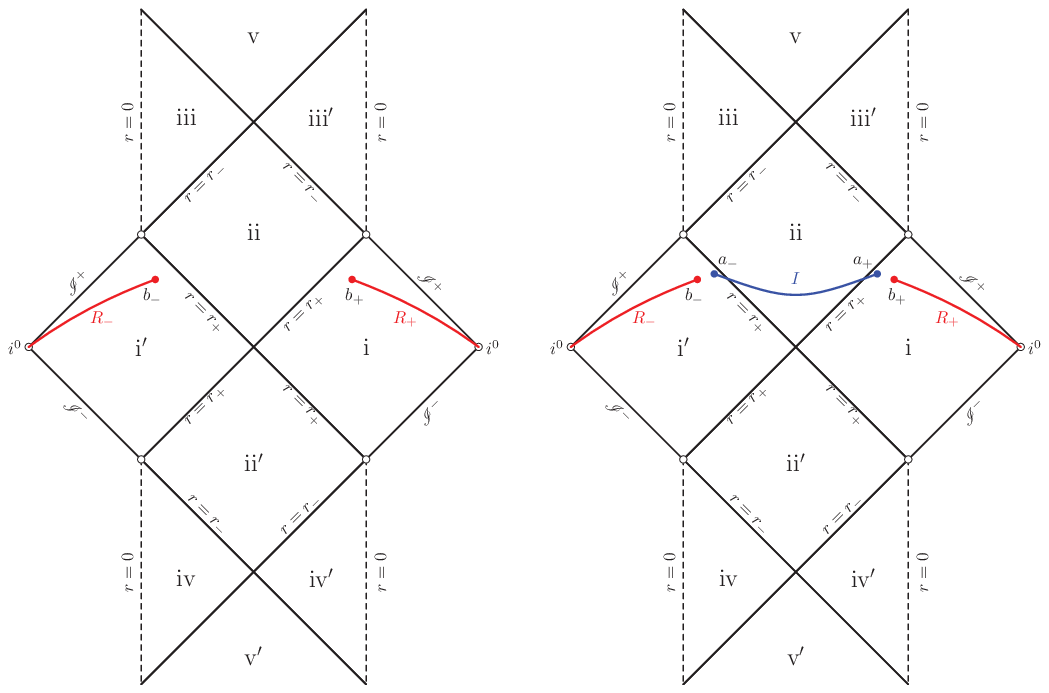} 
	\caption{\label{Fig:PD1}Penrose diagram of the regular black-hole spacetime without an island (left) and that with an island $I$ (right). The radiation region $R$, shown in red, consists of two disconnected parts, $R_+$ and $R_-$, with boundary surfaces at $b_+$ and $b_-$. The island $I$, shown in blue, has boundaries located at $a_+$ and $a_-$. The conformal regions are labeled by i, i$^\prime$, ii, ii$^\prime$, \ldots, and the outer and inner horizons are located at $r=r_+$ and $r=r_-$, respectively.}
\end{figure}
%%%%%%%%%%%%%%%%%%%%%%%%%%%%%%%%%%%%%%%%%%%%%%%%%%%%%%%%%%%%%
\section{Near-horizon geometry of the T-duality black hole}
\label{sec:geometry}
Consider a closed bosonic string in a $(4+1)$-dimensional spacetime with the fifth dimension compactified on a circle of radius $R_5$. The mass spectrum,
\begin{equation}
m^2 = \frac{n^2}{R_5^2} + \frac{w^2 R_5^2}{\alpha'^2} + \frac{2}{\alpha'}\left(N + \tilde N - 2\right),
\end{equation}
is invariant under the T-duality transformation $R_5 \to \alpha'/R_5$, $n \leftrightarrow w$, exchanging Kaluza--Klein momentum modes and winding modes. The self-dual radius $R_5 = \sqrt{\alpha'}$ defines an invariant length below which distances lose operational meaning. An explicit path-integral computation~\cite{Fontanini:2005ik} shows that the
centre-of-mass propagator of such a string, projected to four dimensions, takes the form
\begin{equation}
\mathcal{G}(k) = -\,\frac{\ell_0\; K_1\!\left(\ell_0\sqrt{k^2+m^2}\right)}{\sqrt{k^2+m^2}},
\label{eq:propagator}
\end{equation}
with $\ell_0 = 2\pi\sqrt{\alpha'}$, where $K_1$ is a modified Bessel function of the second kind. For the case $\ell_0\sqrt{k^2+m^2} \ll 1$ one recovers the standard propagator $\sim (k^2+m^2)^{-1}$, while for large momenta the propagator is exponentially damped: the zero-point length acts as an invariant ultraviolet cutoff. Equation \eqref{eq:propagator} coincides with the propagator postulated in path-integral duality \cite{Padmanabhan:1996ap}, confirming that the duality hypothesis is realized in string theory.

The static potential between two masses exchanged by virtual gravitons with propagator \eqref{eq:propagator} is \cite{nicolini2019}
\begin{equation}
V(r) = -\,\frac{M}{\sqrt{r^2 + \ell_0^2}},
\label{eq:potential}
\end{equation}
which is finite at $r=0$ and reduces to the Newtonian potential $-M/r$ for $r \gg \ell_0$. This potential can be interpreted as arising from a smeared source density $\rho(r)$ that is no longer a Dirac delta but a smooth function of width $\ell_0$. Acting with the Laplacian, $\nabla^2 V = 4\pi \rho$, gives the effective smeared density \cite{nicolini2019}
\begin{equation}
\rho(r) = \frac{3 M \ell_0^2}{4\pi (r^2+\ell_0^2)^{5/2}},
\label{eq:smeared}
\end{equation}
whose cumulative mass function is
\begin{equation}
	m(r)=\int_0^{r}4\pi r'^2\rho(r')\,dr'=\frac{M r^{3}}{\left(r^{2}+\ell_0^{2}\right)^{3/2}} .
	\label{eq:mass_function}
\end{equation}
Inserting $m(r)$ into a static, spherically symmetric line element of the form
\begin{equation}
	ds^2=-f(r)\,dt^2+\frac{dr^2}{f(r)}+r^2 d\Omega_2^2 ,
	\label{eq:metric}
\end{equation}
through $f(r)=1-2m(r)/r$ yields the lapse function~\cite{nicolini2019}
\begin{equation}
	f(r)=1-\frac{2Mr^2}{\left(r^2+\ell_0^2\right)^{3/2}} .
	\label{eq:metric_func}
\end{equation}
Here, $\ell_0$ is the minimal length scale associated with the T-duality-inspired regularization, while $M$ denotes the mass of the black hole. The event horizon radius $r_+$ is defined as the largest positive root of $f(r)=0$. Solving this condition for the mass, one obtains
\begin{equation}
	M(r_+)=\frac{\left(r_+^2+\ell_0^2\right)^{3/2}}{2r_+^2}.
	\label{eq:mass}
\end{equation}
The extremal configuration corresponds to the coincidence of the inner and outer horizons, which is encoded in the double-root conditions
\begin{equation}
	\label{eq:extremality-conds}
	f(r_e)=0,\qquad f'(r_e)=0.
\end{equation}
It may also be identified from the minimum of the mass function $M(r_+)$ with respect to $r_+$. These conditions yield
\begin{equation}
	\label{eq:extremal_values}
	r_e=\sqrt{2}\,\ell_0,
	\qquad
	M_{\rm ext}=\frac{3\sqrt{3}}{4}\,\ell_0 .
\end{equation}
For $M>M_{\rm ext}$, the geometry possesses two horizons, while at $M=M_{\rm ext}$ the two horizons coalesce at $r=r_e$.

The Hawking temperature follows from the surface gravity,
$\kappa=f'(r_+)/2$, as
\begin{equation}
	T=\frac{\kappa}{2\pi}
	=\frac{1}{4\pi}f'(r_+)
	=\frac{r_+^2-2\ell_0^2}
	{4\pi r_+\left(r_+^2+\ell_0^2\right)} .
	\label{eq:temp}
\end{equation}
As expected, the temperature vanishes at the extremal radius $r_+=r_e$. The entropy can be obtained by integrating the first law while keeping $\ell_0$ fixed,
\begin{equation}
	dM=T\,dS .
	\label{eq:first_law}
\end{equation}
Using Eq.~\eqref{eq:mass}, one finds
\begin{equation}
	\frac{dS}{dr_+}
	=
	\frac{1}{T}\frac{dM}{dr_+}
	=
	2\pi\,
	\frac{\left(r_+^2+\ell_0^2\right)^{3/2}}{r_+^2}.
	\label{eq:dSdr}
\end{equation}
Integrating with respect to $r_+$ yields
\begin{equation}
	S_{\rm BH}(r_+)
	=
	\pi\,\frac{\left(r_+^2-2\ell_0^2\right)
		\sqrt{r_+^2+\ell_0^2}}{r_+}
	+
	3\pi\ell_0^2
	\log\!\left(
	\frac{r_++\sqrt{r_+^2+\ell_0^2}}{\mu}
	\right),
	\label{eq:entropy}
\end{equation}
where $\mu$ is a reference length scale introduced through the integration constant. In the limit $\ell_0\to0$, Eq.~\eqref{eq:entropy} reduces to the usual area law,
\begin{equation}
	S_{\rm BH}\longrightarrow \pi r_+^2=\frac{A_+}{4},
	\qquad
	A_+=4\pi r_+^2 .
\end{equation}
For finite $\ell_0$, however, the entropy acquires a logarithmic contribution whose precise normalization depends on the choice of the scale $\mu$.

Evaluating Eq.~\eqref{eq:entropy} at the extremal radius, one obtains
\begin{equation}
	S_{\rm ext}
	=
	3\pi\ell_0^2
	\log\!\left(
	\frac{(\sqrt{2}+\sqrt{3})\ell_0}{\mu}
	\right).
	\label{eq:ext_entropy}
\end{equation}
Thus, the extremal remnant carries a finite entropy determined by the minimal length scale $\ell_0$ and by the renormalization convention encoded in $\mu$. Logarithmic terms of this type are familiar from quantum corrections to black hole entropy and often arise in one-loop treatments of the Euclidean path integral \cite{karan2021,sen2012,sen2012a,banerjee2011,banerjee2011a,banerjee2021,iliesiu2025a}. Within the present thermodynamic description, the extremal entropy is therefore entirely accounted for by the logarithmic contribution, while the term that reproduces the standard area law in the limit $\ell_0 \to 0$ vanishes at extremality.

The double-root structure of the extremal configuration motivates the near-horizon parametrization
\begin{equation}
	r=r_e+\rho ,
	\qquad
	|\rho|\ll r_e .
\end{equation}
A Taylor expansion of the lapse function about $r_e$ gives
\begin{equation}
	f(r)
	=
	\frac{1}{2}f''(r_e)\rho^2+\mathcal{O}(\rho^3)
	=
	\frac{2}{3r_e^2}\rho^2+\mathcal{O}(\rho^3).
	\label{eq:f_ext_expansion}
\end{equation}
It is then natural to introduce the length scale $\bar{\ell}_0$ through
\begin{equation}
	f(r)=\frac{\rho^2}{\bar{\ell}_0^2}+\mathcal{O}(\rho^3),
	\qquad
	\bar{\ell}_0^2=\frac{3}{2}r_e^2=3\ell_0^2.
	\label{eq:ads2_radius}
\end{equation}
To leading order in $\rho$, the metric then takes the form
\begin{equation}
	ds^2
	=
	-\frac{\rho^2}{\bar{\ell}_0^2}dt^2
	+
	\frac{\bar{\ell}_0^2}{\rho^2}d\rho^2
	+
	r_e^2 d\Omega_2^2 .
	\label{eq:ext_ads2_metric}
\end{equation}
This is precisely the direct-product geometry $AdS_2\times S^2$, with $AdS_2$ radius $\bar{\ell}_0=\sqrt{3}\,\ell_0$ and sphere radius $r_e=\sqrt{2}\,\ell_0$.

We now consider a small deviation from extremality,
\begin{equation}
	M=M_{\rm ext}+\varepsilon,
	\qquad
	0<\varepsilon\ll M_{\rm ext}.
	\label{eq:near_ext_mass}
\end{equation}
To determine the corresponding near-horizon geometry, we expand the metric function near the extremal radius. This gives
\begin{equation}
	f(r_e+\rho;M_{\rm ext}+\varepsilon)
=
f(r_e;M_{\rm ext})
\,+\,
\left.\frac{\partial f}{\partial M}\right|_{(r_e,M_{\rm ext})}\varepsilon
\,+\,
\left.\frac{\partial f}{\partial r}\right|_{(r_e,M_{\rm ext})}\rho
\,+\,
\frac12 \left.\frac{\partial^2 f}{\partial r^2}\right|_{(r_e,M_{\rm ext})}\rho^2
+\cdots.
	\label{eq:f_near_ext_expansion}
\end{equation}
Using the extremality conditions in Eq.~\eqref{eq:extremality-conds}, the zeroth-order term and the term linear in \(\rho\) vanish, whereas the mass shift \(\varepsilon\) generates the leading deviation from extremality
\begin{equation}
	f(r;M)
	\simeq
	-\frac{4\varepsilon}{3\sqrt{3}\,\ell_0}
	+
	\frac{\rho^2}{\bar{\ell}_0^2}.
	\label{eq:f_near_ext}
\end{equation}
This form suggests introducing a scale $r_0$ that measures the departure from extremality, defined through
\begin{equation}
	f(r;M)\simeq \frac{\rho^2-r_0^2}{\bar{\ell}_0^2}.
	\label{eq:f_near_ext_r0}
\end{equation}
Comparison with Eq.~\eqref{eq:f_near_ext}, together with $\bar{\ell}_0^2=3\ell_0^2$, then yields
\begin{equation}
	r_0^2
	=
	\frac{4\ell_0}{\sqrt{3}}\,\varepsilon .
	\label{eq:r0_epsilon}
\end{equation}
The zeros of $f(r;M)$ are therefore shifted away from the extremal point, and to leading order the two horizons are located at
\begin{equation}
	\rho_\pm=\pm r_0,
	\qquad
	r_\pm=r_e\pm r_0 .
	\label{eq:near_ext_horizons}
\end{equation}
Accordingly, the near-horizon metric takes the form
\begin{equation}
	ds^2
	=
	-\frac{\rho^2-r_0^2}{\bar{\ell}_0^2}\,dt^2
	+
	\frac{\bar{\ell}_0^2}{\rho^2-r_0^2}\,d\rho^2
	+
	r_e^2 d\Omega_2^2 .
	\label{eq:near_ext_metric}
\end{equation}
From this expression, the surface gravity at the outer horizon $\rho=r_0$ follows as
\begin{equation}
	\kappa
	=
	\frac{1}{2}
	\left.
	\frac{d}{d\rho}
	\left(
	\frac{\rho^2-r_0^2}{\bar{\ell}_0^2}
	\right)
	\right|_{\rho=r_0}
	=
	\frac{r_0}{\bar{\ell}_0^2}.
	\label{eq:near_ext_kappa}
\end{equation}
The corresponding near-extremal temperature is therefore
\begin{equation}
	T_{\rm ne}
	=
	\frac{\kappa}{2\pi}
	=
	\frac{r_0}{2\pi \bar{\ell}_0^2}
	=
	\frac{\varepsilon^{1/2}}{3^{5/4} \pi \ell_0^{3/2}}.
	\label{eq:near_ext_temp}
\end{equation}
Thus, the temperature scales as $T_{\rm ne}\propto \sqrt{\varepsilon}$, as expected for a near-extremal black hole. In the following sections, the metric \eqref{eq:near_ext_metric} will be used as the near-horizon background for both the no-island and island analyses.
%%%%%%%%%%%%%%%%%%%%%%%%%%%%%%%%%%%%%%%%%%%%%%%%%%%%%%%%%%%%%
\section{Near-extremal entanglement entropy without islands}
\label{sec:no_island}
We now compute the radiation entropy in the near-horizon region of the near-extremal geometry, in the absence of islands. After dimensional reduction on the $S^2$ factor, the matter sector is described by an effective two-dimensional CFT with central charge $c$, propagating on the $(t,r)$ subspace of the near-horizon metric in Eq.~\eqref{eq:near_ext_metric}. To evaluate the entanglement entropy, we render the two-dimensional line element conformally flat by introducing the Kruskal coordinates
\begin{equation}\label{eq:UV}
	U=-\sqrt{\frac{r-r_e-r_0}{r-r_e+r_0}}
	\exp\!\left(-\frac{r_0}{\bar{\ell}_0^2}t\right),
	\qquad
	V=\sqrt{\frac{r-r_e-r_0}{r-r_e+r_0}}
	\exp\!\left(\frac{r_0}{\bar{\ell}_0^2}t\right).
\end{equation}
In terms of these coordinates, the two-dimensional metric reduces to
\begin{equation}
	ds^2_{(2)}=-\frac{dU\,dV}{W^2},
\end{equation}
where the conformal factor, which carries dimensions of inverse length, is given by
\begin{equation}\label{eq:W}
	W=\frac{r_0}{\bar{\ell}_0}\,\frac{1}{r-r_e+r_0}.
\end{equation}

The radiation region consists of two exterior endpoints located symmetrically on the right and left boundaries, as shown in Fig.~\ref{Fig:PD1}
\begin{equation}\label{eq:bpm}
	b_+:(t,r)=(t_b,b),
	\qquad
	b_-:(t,r)=(-t_b+i\beta/2,b),
\end{equation}
where $\beta=2\pi \bar{\ell}_0^2/r_0$ is the inverse Hawking temperature. The imaginary shift by $i\beta/2$ places the point $b_-$ on the opposite side of the eternal geometry. Using Eqs.~\eqref{eq:no_island_mutual_info} and \eqref{eq:MI_large_sep}, the no-island entropy is computed
from the two-dimensional CFT entropy of the interval with endpoints $b_\pm$. For a conformally
flat metric of the form above, the matter entropy is
\begin{equation}
	S_{\rm matter}
	=
	\frac{c}{6}
	\log\!\left[
	\frac{(U(b_-)-U(b_+))(V(b_+)-V(b_-))}
	{W(b_+)W(b_-)\,\epsilon_{\rm uv}^{2}}
	\right].
	\label{eq:noisland_entropy_general}
\end{equation}
Substituting the near-horizon expressions for $U$, $V$, and $W$ from Eqs.~\eqref{eq:UV} and \eqref{eq:W}, together with the endpoint locations in Eq.~\eqref{eq:bpm}, into Eq.~\eqref{eq:noisland_entropy_general}, we obtain
\begin{equation}
	S_{\rm ne}^{\rm (no\,island)}
	=
	\frac{c}{6}
	\log\!\left[
	\frac{4\bar{\ell}_0^2}{r_0^2}
	\Big((b-r_e)^2-r_0^2\Big)
	\cosh^2\!\left(\frac{r_0 t_b}{\bar{\ell}_0^2}\right)
	\right].
	\label{eq:noisland_final}
\end{equation}
This expression, obtained within the near-horizon approximation, makes the time dependence of the no-island entropy manifest. In particular, it is analogous to the corresponding no-island result for the Schwarzschild black hole \cite{Hashimoto:2020cas},
\begin{equation}
	S_{\rm Sch}^{\rm (no\,island)}=\frac{c}{6} \log \left[\frac{16 r_{\mathrm{h}}^2\left(b-r_{\mathrm{h}}\right)}{b} \cosh ^2 \frac{t_b}{2 r_{\mathrm{h}}}\right].
\end{equation}
The two expressions share the same universal time dependence through a logarithm of $\cosh^2(\kappa t_b)$, where $\kappa$ denotes the corresponding surface gravity. As a result, the first time-dependent correction is quadratic at early times, while the entropy grows linearly at late times,
\begin{align}
	S^{(\mathrm{no\;island})}
	&\simeq
	S_0+\frac{c}{6}\kappa^2 t_b^2
	\qquad
	(t_b\ll \kappa^{-1}),\\
	S^{(\mathrm{no\;island})}
	&\simeq
	S_0+\frac{c}{3}\kappa t_b
	\qquad
	(t_b\gg \kappa^{-1}).
\end{align}
The distinction between the two geometries lies in the scale that sets the surface gravity $\kappa$. 
For Schwarzschild, this is determined by the horizon radius, $\kappa_{\rm Sch} = 1/(2r_h)$, whereas 
for the regular black hole, it is governed by the deviation from extremality, $\kappa = r_0/\bar{\ell}_0^2$. 
Crucially, as the black hole approaches the extremal limit ($r_0 \to 0$), the radiation 
entropy calculated via the no-island prescription in Eq.~\eqref{eq:noisland_final} 
exhibits a logarithmic divergence proportional to $\log(r_0^{-2})$. This divergence signifies the breakdown of the semi-classical no-island approximation 
as the system nears extremality, effectively manifesting a near-horizon version of the 
information paradox. Consequently, the island prescription is a fundamental requirement to regulate this divergence. By introducing the island, 
the generalized entropy remains finite, thereby ensuring a physically consistent Page curve 
that naturally terminates the entropy growth as $r_0 \to 0$. This resolution 
distinguishes the regular black hole from the Schwarzschild case, where the 
divergence is driven by the evaporation to zero mass, necessitating a different 
mechanism for the final stage of the Page curve.

%%%%%%%%%%%%%%%%%%%%%%%%%%%%%%%%%%%%%%%%%%%%%%%%%%%%%%%%%%%%%
\section{Island phase with the area functional}
\label{sec:island}
In this section, we evaluate the generalized entropy using the island prescription,
Eq.~\eqref{eq:island}. The corresponding functional comprises two distinct contributions:
(i) the gravitational area term associated with the island boundary and (ii) the von Neumann
entropy of bulk matter fields on $R\cup I$. We work in the near-horizon regime in which the
island endpoints $a_\pm$ and the radiation endpoints $b_\pm$ lie close to the outer horizon
at $r=r_+$, as shown in the right panel of Fig.~\ref{Fig:PD1}. In particular, the separations
between the paired endpoints $(a_\pm,b_\pm)$ are taken to be much smaller than the
characteristic curvature scale $r_e$. The matter entropy therefore reduces to the
short-interval expression given in Eq.~\eqref{eq:Information-Area}. Let us denote the coordinates of the island boundaries $a_\pm$ as $(t, r) = (t_a, a)$ for 
$a_+$ and $(t, r) = (-t_a + i\beta/2, a)$ for $a_-$. Similarly, the radiation boundaries 
$b_\pm$ are located at $(t_b, b)$ and $(-t_b + i\beta/2, b)$, respectively. To extremize the 
generalized entropy, it is plausible to assume $t_a = t_b$. Furthermore, in the late-time regime, the left and right wedges are separated by an
interior region whose volume grows linearly with time
\cite{jiang2025,Hashimoto:2020cas}. Therefore, it is sufficient to perform the entropy
calculation on the right-hand side of the Penrose diagram. The contribution from the
left-hand side is identical by symmetry, and hence the final generalized entropy is
obtained by doubling the right-wedge result. Accordingly, restricting attention to the right wedge of the Penrose diagram, we denote
the island and radiation endpoints $a_+$ and $b_+$ simply by $a$ and $b$, respectively. The proper radial distance between these two endpoints is then given by
\begin{equation}
	L(a,b)
	=
	\int_a^b \frac{dr}{\sqrt{f(r)}} =\bar{\ell}_0 \log\!\left(
	\frac{b-r_e+\sqrt{(b-r_e)^2-r_0^2}}
	{a-r_e+\sqrt{(a-r_e)^2-r_0^2}}
	\right).
\end{equation}
The generalized entropy of the radiation region including the island is therefore
\begin{equation}
	S_{\rm gen}(a)
	=
	\pi a^2
	-
	\frac{\pi\lambda\, b^2}{\left(\log\left[
		\frac{b-r_e+\sqrt{(b-r_e)^2-r_0^2}}
		{a-r_e+\sqrt{(a-r_e)^2-r_0^2}}
		\right]\right)^2},
	\label{eq:Sgen-classical}
\end{equation}
with $\lambda$ defined in Eq.~\eqref{eq:lambda_def}.
The location of the island endpoint, $a=\widetilde{a}$, is determined by
extremizing the generalized entropy with respect to $a$, which yields
\begin{equation}
	\label{eq:extremization-classical}
	\widetilde{a} \sqrt{(\widetilde{a}-r_e)^2-r_0^2} \left(\log \left[\frac{b-r_e+\sqrt{(b-r_e)^2-r_0^2}}{\widetilde{a}-r_e+\sqrt{(\widetilde{a}-r_e)^2-r_0^2}}\right]\right)^3=\frac{4 \chi c}{\bar{\ell}_0^2} \,b^2\;=\;\lambda\, b^2 .
\end{equation}
It is convenient to introduce the dimensionless parameters $\delta_a$ and $\delta_b$ via
\begin{equation}
	a-r_e-r_0= \delta_a r_0,
	\qquad
	b-r_e-r_0= \delta_b r_0,
	\qquad
	r_0\ll r_e.
\end{equation}
In the near-horizon, near-extremal regime $r_0/r_e\ll 1$, the parameters
$\delta_a$ and $\delta_b$ are kept finite. Expressed in terms of these
dimensionless variables, the extremization condition becomes
\begin{align}
	\label{eq:extremality-deltas}
	\left(1+\left(\widetilde{\delta_a}+1\right)\frac{r_0}{r_e}\right)\sqrt{\widetilde{\delta_a}(\widetilde{\delta_a}+2)} &\left(\log \left[\frac{1+\delta_b+\sqrt{\delta_b(\delta_b+2)}}{1+\widetilde{\delta_a}+\sqrt{\widetilde{\delta_a}(\widetilde{\delta_a}+2)}}\right]\right)^3
	\\\nonumber
	&=\lambda\,\frac{r_e}{r_0}\left(1+2\left(\delta_b+1\right)\frac{r_0}{r_e}+\left(\delta_b+1\right)^2\frac{r_0^2}{r_e^2}\right),
\end{align}
where $\widetilde{\delta}_a$ denotes the extremizing value of $\delta_a$. The condition is
highly nonlinear and does not admit a simple closed-form solution; we therefore solve it
numerically, using the representative value $\lambda=10^{-2}$ and the near-extremal ratios
$r_e/r_0=10,10^{2},10^{3}$.

Before turning to the numerics it is useful to establish which of the solutions is
physical. As $a\to r_+$ the radial proper distance to the radiation boundary approaches
its maximum value at fixed $b$, while $\sqrt{(a-r_e)^2-r_0^2}\to0$, so that
\begin{equation}
\frac{dS_{\rm gen}}{da}
=2\pi a-\frac{2\pi\lambda\, b^{2}}
{\sqrt{(a-r_e)^{2}-r_0^{2}}\;\Lambda(a)^{3}}
\;\longrightarrow\;-\infty ,
\qquad a\to r_+^{\,+},
\label{eq:dSgen_horizon}
\end{equation}
where $\Lambda(a)$ denotes the logarithm appearing in Eq.~\eqref{eq:Sgen-classical}. The
generalized entropy therefore \emph{decreases} as the island boundary is displaced
outwards from the horizon, and the first stationary point encountered is necessarily a
local minimum. At the second stationary point $S_{\rm gen}$ turns over into a local
maximum, beyond which it decreases without bound as $a\to b$; this runaway is spurious and
merely reflects the breakdown of the short-distance form \eqref{eq:Information-Area} when
$L\to0$, where the mutual information is no longer bounded by the physical entropy. The
physical island is thus the \emph{smaller} of the two roots.

\begin{figure}[t]
	\centering
	\includegraphics[width=0.72\textwidth]{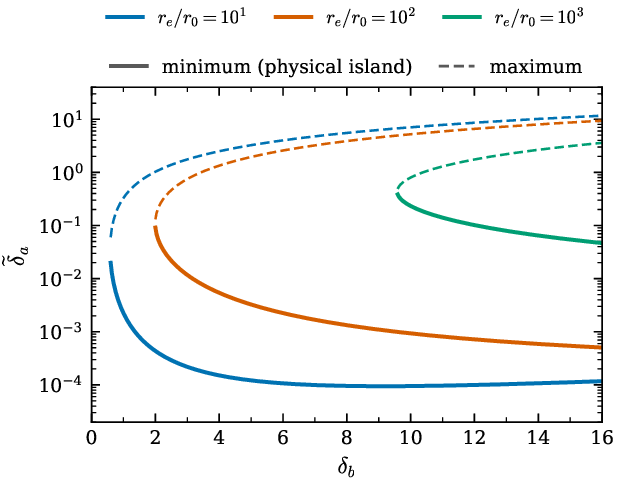}
	\caption{\label{Fig:deltas}Stationary points $\widetilde{\delta}_a$ of the generalized
		entropy as a function of $\delta_b$, for $\lambda=10^{-2}$ and the three
		representative near-extremal ratios $r_e/r_0=10$ (blue), $10^{2}$ (orange) and
		$10^{3}$ (green). Solid curves are the minima of $S_{\rm gen}$, and hence the
		physical islands; dashed curves are the maxima. Each pair terminates on the left
		at the threshold value of $\delta_b$ below which no island exists; the threshold
		grows as extremality is approached, which is why the $r_e/r_0=10^{3}$ pair appears
		only beyond $\delta_b\simeq9.6$. Note the logarithmic vertical scale: the physical
		branch lies several orders of magnitude closer to the horizon than the spurious
		one, and the gap widens towards extremality.}
\end{figure}

Fig.~\ref{Fig:deltas} displays both branches. For each near-extremal ratio the solid curve
is the minimum and the dashed curve the maximum. Islands exist only above a threshold value
of $\delta_b$: the right-hand side of Eq.~\eqref{eq:extremality-deltas} grows as
$r_e/r_0$, so a nearer-extremal black hole requires the radiation boundary to be placed
further out before an island forms. For $\lambda=10^{-2}$ this threshold is
$\delta_b\simeq0.58,\,2.0$ and $9.6$ for $r_e/r_0=10,10^{2}$ and $10^{3}$ respectively.

On the physical branch the island lies extremely close to the outer horizon and an
analytic solution is available. For $\widetilde\delta_a\ll1$ one has
$\sqrt{\widetilde\delta_a(\widetilde\delta_a+2)}\simeq\sqrt{2\widetilde\delta_a}$, while the
logarithm saturates at its horizon value
\begin{equation}
\Lambda_b\;\equiv\;\log\!\left[1+\delta_b+\sqrt{\delta_b(\delta_b+2)}\right],
\label{eq:Lambda_b}
\end{equation}
so that Eq.~\eqref{eq:extremality-deltas} collapses to
\begin{equation}
\widetilde{a}-r_+\;\simeq\;\frac{\lambda^{2}\,b^{4}}{2\,r_e^{2}\,r_0\,\Lambda_b^{6}} .
\label{eq:island_analytic_classical}
\end{equation}
For $\lambda=10^{-2}$, $b=1.10\,r_e$ and $r_0/r_e=10^{-2}$ this gives
$(\widetilde a-r_+)/r_e=1.02\times10^{-5}$ against the numerical value
$1.10\times10^{-5}$, the residual difference coming from the
$\mathcal{O}(\sqrt{\widetilde\delta_a}/\Lambda_b)$ correction to the logarithm. Eq.~\eqref{eq:island_analytic_classical} makes the scaling
transparent: the displacement is second order in the semiclassical coupling and is
suppressed by the sixth power of the proper separation, which is why the island is pinned
so tightly to the horizon.

\begin{figure}[t]
	\centering
	\includegraphics[width=0.72\textwidth]{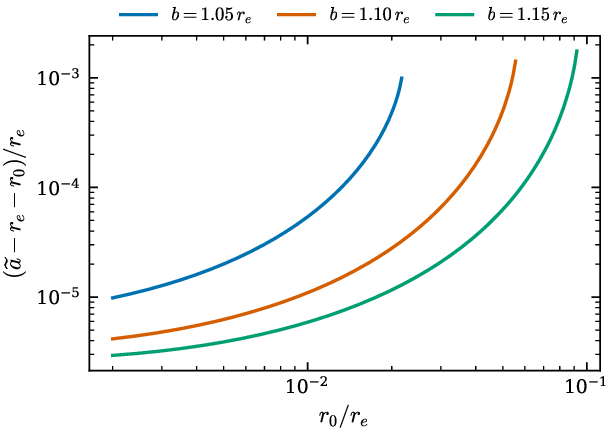}
	\caption{The displacement of the island boundary from the outer horizon,
		$(\widetilde{a}-r_e-r_0)/r_e$, as a function of the extremality parameter
		$r_0/r_e$, for $b=1.05\,r_e$ (blue), $1.10\,r_e$ (orange) and $1.15\,r_e$
		(green), at $\lambda=10^{-2}$. Both axes are logarithmic. The displacement never
		exceeds $\sim10^{-3}r_e$ over the range shown and decreases monotonically towards
		extremality, so that the island stays pinned to the immediate vicinity of the
		outer horizon. Each curve terminates at the largest $r_0$ for which an island
		solution exists at that value of $b$.}
	\label{Fig:atilde}
\end{figure}

Fig.~\ref{Fig:atilde} confirms this over the whole near-extremal window. For the three
radiation endpoints $b/r_e=1.05,1.10,1.15$ the displacement never exceeds
$\sim10^{-3}r_e$ and falls below $10^{-5}r_e$ as extremality is approached. This is in
good agreement with the Schwarzschild analysis of Ref.~\cite{Hashimoto:2020cas}, where the
island surface is likewise pinned to the near-horizon region. The monotonic decrease
towards extremality follows directly from
Eq.~\eqref{eq:island_analytic_classical}: although the explicit factor $r_0^{-1}$ grows,
$\delta_b=(b-r_e)/r_0-1$ grows faster, and the resulting sixth power of
$\Lambda_b\simeq\log(2(b-r_e)/r_0)$ wins over the entire range shown.

\begin{figure}[t]
	\centering
	\includegraphics[width=0.72\textwidth]{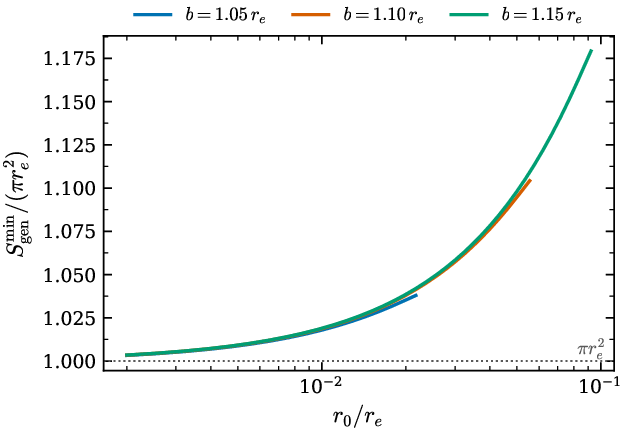}
	\caption{The minimized generalized entropy, normalized by the extremal area
		$\pi r_e^2$, as a function of the extremality parameter $r_0/r_e$ for
		$b=1.05\,r_e$ (blue), $1.10\,r_e$ (orange) and $1.15\,r_e$ (green), at
		$\lambda=10^{-2}$. In the extremal limit the entropy approaches the extremal
		horizon area, Eq.~\eqref{eq:plateau_classical}, shown as the dotted line. The
		three curves are nearly indistinguishable because the matter correction enters
		only through $\Lambda_b^{-2}$, which depends on $b$ only logarithmically.}
	\label{Fig:Sgen}
\end{figure}

Finally we evaluate the generalized entropy at the minimum. Because
$\widetilde{a}-r_+$ is of order $\lambda^{2}$, the area term may be evaluated on the
horizon itself and
\begin{equation}
S_{\rm gen}^{\rm min}
\;\simeq\;
\pi r_+^{2}-\frac{\pi\lambda\,b^{2}}{\Lambda_b^{2}}
+\mathcal{O}(\lambda^{2}),
\label{eq:Sgen_min_classical}
\end{equation}
i.e.\ the Bekenstein--Hawking entropy of the outer horizon, reduced by the mutual
information between the island and the radiation. Fig.~\ref{Fig:Sgen} shows this quantity
normalized by the extremal horizon area. As the black hole approaches extremality the
matter correction becomes negligible --- it is suppressed by $\Lambda_b^{-2}$, which grows
logarithmically as $r_0\to0$ --- and the minimized generalized entropy approaches the
extremal Bekenstein--Hawking value,
\begin{equation}
\lim_{r_0\to0}S_{\rm gen}^{\rm min}=\pi r_e^{2}=\frac{A_{\rm ext}}{4}.
\label{eq:plateau_classical}
\end{equation}
The plateau of the Page curve in this prescription is therefore exactly the extremal
horizon area. As we discuss in Sec.~\ref{sec:island_quantum}, this is in tension with the
thermodynamic result \eqref{eq:ext_entropy}, which states that the area contribution to the
entropy of the extremal configuration vanishes identically.

\section{Island with the quantum-corrected entropy functional}
\label{sec:island_quantum}

In the previous section the gravitational contribution to the generalized entropy was
taken to be the bare area term $\mathrm{Area}(\partial I)/4=\pi a^{2}$. This is the
standard choice, but it is not the only consistent one. The island prescription in the
form of Eq.~\eqref{eq:renormalized_prescription} instructs us to extremize the
\emph{total} gravitational entropy associated with the entangling surface. Whenever the
theory assigns to a sphere of radius $r$ an entropy that differs from $\pi r^{2}$ --- as
happens here, where integration of the first law produced the logarithmically corrected
expression \eqref{eq:entropy} --- it is natural to ask whether the island functional
should be built from that same entropy. In this section we explore this possibility
systematically: we replace
\begin{equation}
\pi a^{2}\;\longrightarrow\;S_{\rm BH}(a)
=\pi\,\frac{(a^{2}-2\ell_0^{2})\sqrt{a^{2}+\ell_0^{2}}}{a}
+3\pi\ell_0^{2}\log\!\left(\frac{a+\sqrt{a^{2}+\ell_0^{2}}}{\mu}\right),
\label{eq:SBH_of_a}
\end{equation}
i.e.\ we evaluate Eq.~\eqref{eq:entropy} on the island boundary by the substitution
$r_+\to a$. Since $S_{\rm BH}(r_+)\to\pi r_+^{2}$ as $\ell_0\to 0$, the classical
analysis of Sec.~\ref{sec:island} is recovered smoothly in that limit, and the two
prescriptions differ only through the minimal length scale.

\subsection{The modified generalized entropy}

With the same short-distance matter contribution as before, the generalized entropy in
the right wedge reads
\begin{equation}
S_{\rm gen}^{\rm (q)}(a)
=S_{\rm BH}(a)
-\frac{\pi\lambda\,b^{2}}
{\left(\log\!\left[
\dfrac{b-r_e+\sqrt{(b-r_e)^{2}-r_0^{2}}}
{a-r_e+\sqrt{(a-r_e)^{2}-r_0^{2}}}\right]\right)^{2}} .
\label{eq:Sgen_quantum}
\end{equation}
The essential simplification is that the derivative of \eqref{eq:SBH_of_a} is
\emph{algebraic}: using Eq.~\eqref{eq:dSdr} with $r_+\to a$,
\begin{equation}
\frac{dS_{\rm BH}}{da}=2\pi\,\Phi(a),
\qquad
\Phi(a)\equiv\frac{\left(a^{2}+\ell_0^{2}\right)^{3/2}}{a^{2}} .
\label{eq:Phi_def}
\end{equation}
The function $\Phi(a)$ therefore plays the role of an \emph{effective radius}: every step
of the classical derivation goes through verbatim with $a\to\Phi(a)$ in the area
variation. Extremizing \eqref{eq:Sgen_quantum} with respect to $a$ gives
\begin{equation}
\Phi(\widetilde{a})\,
\sqrt{(\widetilde{a}-r_e)^{2}-r_0^{2}}
\left(\log\!\left[\frac{b-r_e+\sqrt{(b-r_e)^{2}-r_0^{2}}}
{\widetilde{a}-r_e+\sqrt{(\widetilde{a}-r_e)^{2}-r_0^{2}}}\right]\right)^{3}
=\lambda\,b^{2},
\label{eq:extremality_quantum}
\end{equation}
which is Eq.~\eqref{eq:extremality-deltas} with the single replacement
$\widetilde{a}\to\Phi(\widetilde{a})$. Note that the renormalization scale $\mu$ enters
\eqref{eq:Sgen_quantum} only as an additive constant, so the island location is
completely independent of $\mu$; the scheme dependence affects the \emph{value} of the
entropy but never its extremum.

\subsection{Near-extremal reduction: an effective coupling}

The key structural fact is that $\Phi$ is stationary exactly at the extremal radius,
\begin{equation}
\Phi'(a)=\frac{\sqrt{a^{2}+\ell_0^{2}}\,\left(a^{2}-2\ell_0^{2}\right)}{a^{3}},
\qquad
\Phi'(r_e)=0,
\qquad
r_e=\sqrt{2}\,\ell_0 ,
\label{eq:Phi_prime}
\end{equation}
because the vanishing of $a^{2}-2\ell_0^{2}$ is the extremality condition itself. Writing
$a=r_e(1+u)$ with $u=(1+\delta_a)\,r_0/r_e\ll1$ one finds
\begin{equation}
\frac{\Phi(a)}{r_e}
=\left(\frac{3}{2}\right)^{3/2}\left[1+\frac{2}{3}u^{2}+\mathcal{O}(u^{3})\right]
=\frac{3\sqrt{6}}{4}\left[1+\mathcal{O}\!\left(\frac{r_0^{2}}{r_e^{2}}\right)\right].
\label{eq:Phi_expansion}
\end{equation}
The linear term is absent, so throughout the near-horizon, near-extremal window the
corrected area weight is a \emph{constant} enhancement of the classical one by the
factor $(3/2)^{3/2}\simeq1.837$. In terms of the dimensionless variables
$\delta_a,\delta_b$ the extremality condition therefore becomes
\begin{equation}
\left(\frac{3}{2}\right)^{3/2}
\sqrt{\widetilde{\delta_a}(\widetilde{\delta_a}+2)}
\left(\log\!\left[\frac{1+\delta_b+\sqrt{\delta_b(\delta_b+2)}}
{1+\widetilde{\delta_a}+\sqrt{\widetilde{\delta_a}(\widetilde{\delta_a}+2)}}\right]\right)^{3}
=\lambda\,\frac{r_e}{r_0}
\left(1+(\delta_b+1)\frac{r_0}{r_e}\right)^{2},
\label{eq:extremality_deltas_quantum}
\end{equation}
up to corrections of relative order $(r_0/r_e)^{2}$. Comparing with
Eq.~\eqref{eq:extremality-deltas}, the entire effect of the logarithmic correction in the
near-extremal regime is the rescaling of the dimensionless coupling
\begin{equation}
\lambda\;\longrightarrow\;
\lambda_{\rm eff}=\left(\frac{2}{3}\right)^{3/2}\lambda\simeq0.544\,\lambda .
\label{eq:lambda_eff}
\end{equation}
Solving \eqref{eq:extremality_deltas_quantum} numerically and solving
\eqref{eq:extremality-deltas} with $\lambda\to\lambda_{\rm eff}$ agree to better than
$1.5\%$ at $r_0/r_e=10^{-2}$, the residual difference being the $\mathcal{O}(r_0/r_e)$
term retained on the classical side.

\subsection{Location of the island}

The branch analysis of Sec.~\ref{sec:island} carries over unchanged: since
$\Phi(a)$ is finite and positive, the argument leading to Eq.~\eqref{eq:dSgen_horizon}
is unaffected, the smaller root of \eqref{eq:extremality_quantum} remains the minimum, and
the island again lies parametrically close to the outer horizon.

\begin{figure}[t]
\centering
\includegraphics[width=0.72\textwidth]{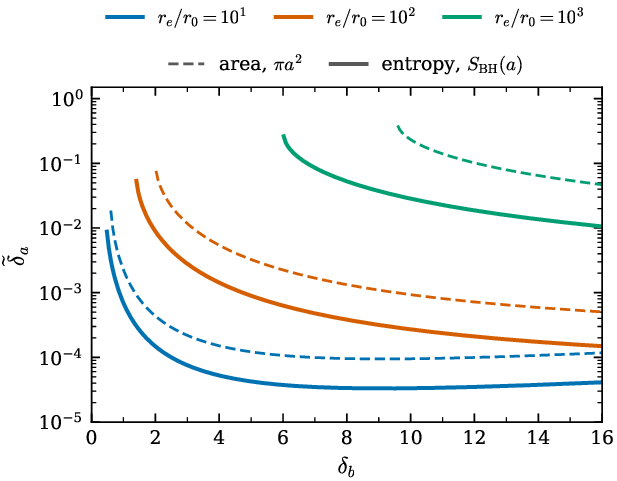}
\caption{\label{Fig:deltas_q}Position of the physical island, $\widetilde{\delta}_a$,
as a function of $\delta_b$, for $\lambda=10^{-2}$ and $r_e/r_0=10$ (blue), $10^{2}$
(orange), $10^{3}$ (green). Dashed curves use the area functional $\pi a^{2}$, solid
curves the corrected functional $S_{\rm BH}(a)$. Each curve begins at the threshold in
$\delta_b$ above which an island exists; the corrected functional lowers that threshold,
which is why the solid curves extend further to the left than the dashed ones of the same
colour. It also places the island closer to the horizon, by the constant factor $8/27$ of
Eq.~\eqref{eq:ratio_8_27} once $\widetilde\delta_a\ll1$.}
\end{figure}

\begin{figure}[t]
\centering
\includegraphics[width=0.72\textwidth]{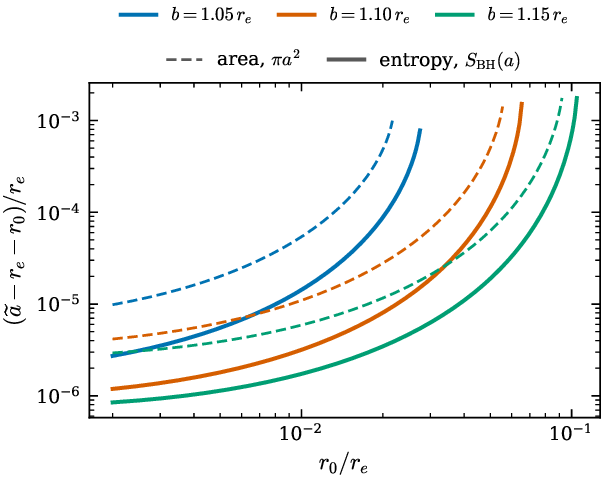}
\caption{\label{Fig:a_q}Displacement of the island boundary from the outer horizon,
$(\widetilde{a}-r_e-r_0)/r_e$, versus $r_0/r_e$, for $b=1.05\,r_e$ (blue), $1.10\,r_e$
(orange) and $1.15\,r_e$ (green), at $\lambda=10^{-2}$. Dashed curves use the area
functional, solid curves the corrected entropy functional. In both cases the island
remains pinned to the immediate vicinity of $r_+$; the corrected functional reduces the
displacement by the constant factor $8/27$ of Eq.~\eqref{eq:ratio_8_27}, visible here as
a rigid vertical offset of the solid curves relative to the dashed ones.}
\end{figure}

Repeating the analysis that led to Eq.~\eqref{eq:island_analytic_classical}, with $r_e$
replaced by $\Phi(r_e)=(3/2)^{3/2}r_e$, gives
\begin{equation}
\widetilde{a}-r_+
\;\simeq\;
\frac{1}{2\,r_0\,\Lambda_b^{6}}
\left(\frac{\lambda\,b^{2}}{\Phi(r_e)}\right)^{2}
=\left(\frac{2}{3}\right)^{3}
\frac{\lambda^{2}b^{4}}{2\,r_e^{2}\,r_0\,\Lambda_b^{6}} ,
\label{eq:island_analytic}
\end{equation}
with $\Lambda_b$ as in Eq.~\eqref{eq:Lambda_b}; the classical result is recovered by
dropping the prefactor $(2/3)^{3}$. The corrected functional therefore pushes the island \emph{closer} to the horizon by the
universal factor
\begin{equation}
\frac{\widetilde{a}^{\,\rm (q)}-r_+}{\widetilde{a}^{\,\rm (cl)}-r_+}
\;\longrightarrow\;\left(\frac{2}{3}\right)^{3}=\frac{8}{27}\simeq0.296 .
\label{eq:ratio_8_27}
\end{equation}
Numerically, for $\lambda=0.01$, $b=1.10\,r_e$ and $r_0/r_e=10^{-2}$ we obtain
$(\widetilde{a}-r_+)/r_e=1.10\times10^{-5}$ classically against
$3.17\times10^{-6}$ with \eqref{eq:SBH_of_a}, i.e.\ a ratio $0.289$, in good agreement
with \eqref{eq:ratio_8_27}. Fig.~\ref{Fig:deltas_q} shows both branches of
$\widetilde{\delta}_a(\delta_b)$ for the two prescriptions, and Fig.~\ref{Fig:a_q}
displays the displacement of the island from the outer horizon as a function of the
extremality parameter.

\subsection{Saturation value and the extremal plateau}

\begin{figure}[t]
\centering
\includegraphics[width=0.98\textwidth]{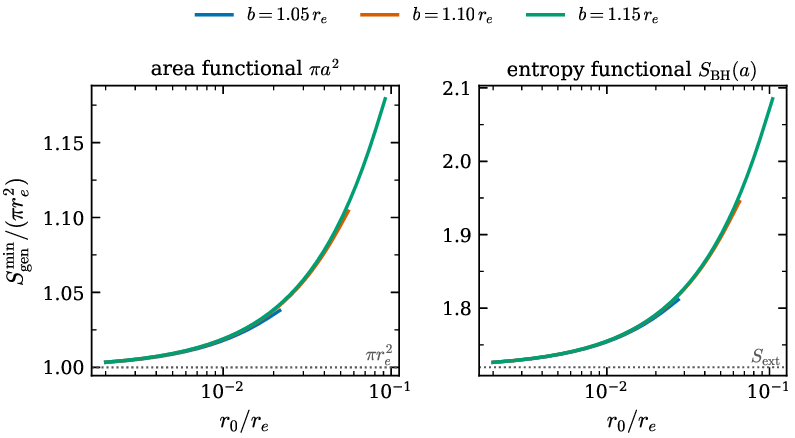}
\caption{\label{Fig:Sgen_q}Minimized generalized entropy normalized by $\pi r_e^{2}$ as
a function of $r_0/r_e$ for $b=1.05\,r_e$ (blue), $1.10\,r_e$ (orange) and $1.15\,r_e$
(green), at $\lambda=10^{-2}$. Left: area functional, which approaches the extremal
horizon area $\pi r_e^{2}$. Right: corrected entropy functional, which approaches the
purely logarithmic extremal entropy $S_{\rm ext}$ of Eq.~\eqref{eq:ext_entropy}, shown
here for the representative choice $\mu=\ell_0$, for which
$S_{\rm ext}/\pi r_e^{2}=\tfrac32\log(\sqrt{2}+\sqrt{3})\simeq1.719$. Dotted lines mark
the two extremal limits. Note the different vertical scales of the two panels: the
plateau of the Page curve differs by roughly a factor of two between the two
prescriptions.}
\end{figure}

Because the island sits on the horizon to leading order, the minimized generalized
entropy is obtained by evaluating \eqref{eq:Sgen_quantum} at $a=r_+$,
\begin{equation}
S_{\rm gen}^{\rm (q),min}
\;\simeq\;
S_{\rm BH}(r_+)-\frac{\pi\lambda\,b^{2}}{\Lambda_b^{2}}
+\mathcal{O}\!\left(\lambda^{2}\right),
\label{eq:Sgen_min_quantum}
\end{equation}
to be compared with Eq.~\eqref{eq:Sgen_min_classical}.
This is the central physical difference between the two prescriptions. The Page curve now
saturates not at the horizon area but at the full quantum-corrected entropy, and using
$S_{\rm BH}(r_e+r_0)=S_{\rm ext}+3\sqrt{3}\pi\ell_0 r_0+\mathcal{O}(r_0^{3})$ --- the
quadratic term is absent precisely because of \eqref{eq:Phi_prime} --- the extremal limit
of the plateau is
\begin{equation}
\lim_{r_0\to0}S_{\rm gen}^{\rm (q),min}
=S_{\rm ext}
=3\pi\ell_0^{2}\log\!\left(\frac{(\sqrt{2}+\sqrt{3})\,\ell_0}{\mu}\right),
\label{eq:plateau_extremal}
\end{equation}
whereas the classical functional gives $\lim_{r_0\to0}S_{\rm gen}^{\rm (cl),min}=\pi r_e^{2}$.
The corrected prescription is thus internally consistent with the thermodynamic statement
of Sec.~\ref{sec:geometry}: at extremality the area-law piece vanishes identically and the
entire entropy --- and hence the entire information-storage capacity of the remnant --- is
carried by the logarithmic term. In the classical treatment the two statements are in
tension, since the island plateau retains a finite area contribution $\pi r_e^{2}$ that the
first law says is absent. Fig.~\ref{Fig:Sgen_q} shows the two saturation curves.

\subsection{Remarks and caveats}

Three comments are in order. First, the value of the plateau \eqref{eq:plateau_extremal} inherits the renormalization
ambiguity of $\mu$ and can in principle be made arbitrarily small, or negative, for
$\mu>(\sqrt{2}+\sqrt{3})\ell_0$. As emphasized above, $\mu$ does not affect the island
location, so the geometry of the island phase is unambiguous; only the height of the Page
plateau is scheme dependent. A physically motivated choice is to fix $\mu$ by matching to
the microcanonical count of states of the extremal remnant, or simply to set
$\mu\sim\ell_0$, in which case $S_{\rm ext}=\tfrac32\pi r_e^{2}\log(\sqrt2+\sqrt3)$ is of
the order of the extremal horizon area. There is in fact a sharp upper bound on $\mu$,
and a distinguished value at which it is saturated; we return to this in
Sec.~\ref{sec:mu}.

Second, the substitution $r_+\to a$ in \eqref{eq:SBH_of_a} is an assumption, not a
theorem. Equation~\eqref{eq:entropy} was obtained by integrating the first law
\emph{on the horizon}, and its continuation to an arbitrary sphere at $r=a$ is justified
only if the logarithmic term descends from a local, covariant entropy functional --- a
Wald-like contribution --- rather than from a
global one-loop determinant. The clean structure found above, in particular the fact that
$dS_{\rm BH}/da$ is the simple algebraic function \eqref{eq:Phi_def}, is suggestive but
not decisive. It would be worth checking whether the T-duality-inspired action of
Ref.~\cite{nicolini2019} admits such a local functional; if it does, the prescription
used in this section is the correct one, and the classical area functional of
Sec.~\ref{sec:island} should be regarded as its $\ell_0\to0$ approximation.

Third, both prescriptions leave the no-island phase of Sec.~\ref{sec:no_island}
untouched, since in that configuration no entangling surface sits in the gravitating
region. The two prescriptions therefore differ only through the height of the plateau,
and hence through the Page time; this is worked out in the next section.

%%%%%%%%%%%%%%%%%%%%%%%%%%%%%%%%%%%%%%%%%%%%%%%%%%%%%%%%%%%%%
\section{The Page curve}
\label{sec:page}

The entropy of the Hawking radiation is the smaller of the two competing saddles,
\begin{equation}
S(R)=\min\left\{\,S^{\rm (no\;island)}_{\rm ne}(t_b)\;,\;S_{\rm gen}^{\rm min}\,\right\},
\label{eq:page_min}
\end{equation}
with the time-dependent no-island branch given by Eq.~\eqref{eq:noisland_final} and the
constant island branch by Eq.~\eqref{eq:Sgen_min_classical} or
\eqref{eq:Sgen_min_quantum} according to the prescription adopted. At early times the
no-island configuration dominates and the entropy grows, linearly for
$t_b\gg\kappa^{-1}$; at the Page time the island configuration takes over and the entropy
saturates. Equating the two branches gives
\begin{equation}
t_{\rm Page}
\simeq
\frac{\bar{\ell}_0^{2}}{r_0}
\left[\frac{3}{c}\,S_{\rm gen}^{\rm min}
-\frac{1}{2}\log\!\left(\frac{4\bar{\ell}_0^{2}\big((b-r_e)^{2}-r_0^{2}\big)}{r_0^{2}}\right)\right].
\label{eq:page_time}
\end{equation}
The first term dominates whenever the plateau is large in units of the central charge,
which is the semiclassical regime of interest.

The two prescriptions differ only through $S_{\rm gen}^{\rm min}$, and hence
\begin{equation}
\Delta t_{\rm Page}
=t_{\rm Page}^{\rm (q)}-t_{\rm Page}^{\rm (cl)}
\simeq\frac{3\bar{\ell}_0^{2}}{c\,r_0}\Big[S_{\rm BH}(r_+)-\pi r_+^{2}\Big].
\label{eq:delta_page_time}
\end{equation}
For the natural choice $\mu\lesssim\ell_0$ one has $S_{\rm BH}(r_+)>\pi r_+^{2}$, so the
logarithmic correction \emph{delays} the Page transition. Moreover the delay grows without
bound as $r_0\to0$: in the extremal limit the bracket tends to
$S_{\rm ext}-\pi r_e^{2}$, a finite number, while the prefactor $\bar{\ell}_0^{2}/r_0$
diverges. The near-extremal black hole therefore radiates for parametrically longer before
its radiation entropy saturates, consistent with the long-lived remnant picture developed
in Sec.~\ref{sec:discussion}.

\begin{figure}[t]
	\centering
	\includegraphics[width=0.78\textwidth]{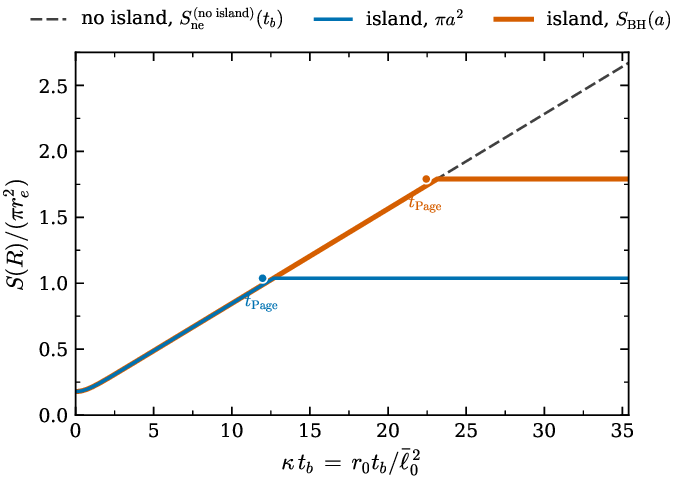}
	\caption{\label{Fig:Page-Curve}Page curve of the Hawking radiation for
		$r_0/r_e=0.02$, $b=1.10\,r_e$, $c=100$, $\lambda=10^{-2}$ and $\mu=\ell_0$.
		The dashed grey line is the no-island entropy \eqref{eq:noisland_final}, which
		grows linearly at late times. The blue curve uses the area functional $\pi a^{2}$
		and saturates at $S_{\rm gen}^{\rm min}\simeq1.04\,\pi r_e^{2}$; the orange curve
		uses the corrected functional $S_{\rm BH}(a)$ and saturates at
		$1.79\,\pi r_e^{2}$. Dots mark the corresponding Page times,
		$\kappa t_{\rm Page}=12.0$ and $22.4$. Time is measured in units of the inverse
		surface gravity $\kappa^{-1}=\bar{\ell}_0^{2}/r_0$. Both prescriptions give a
		Page curve of the standard shape; they differ in the height of the plateau and
		hence in the time at which it is reached.}
\end{figure}

Fig.~\ref{Fig:Page-Curve} shows the resulting Page curves for a representative
near-extremal configuration. Both prescriptions produce a curve of the standard shape ---
linear growth followed by an exact plateau --- so unitarity is restored in either case;
what differs is the height of the plateau and the time at which it is reached.

%\subsection{Why the plateau is flat: the eternal approximation}
%\label{sec:eternal}

The Page curve of Fig.~\ref{Fig:Page-Curve} rises and then saturates, but never turns
over. This is not an artefact: it is the correct behaviour for the configuration we have
analysed. Throughout Secs.~\ref{sec:no_island}--\ref{sec:island_quantum} the geometry is
the \emph{eternal} two-sided solution, held in equilibrium with the bath, so that the
outgoing Hawking flux is exactly balanced by an ingoing one. The metric functions
$f(r)$, the surface gravity $\kappa$ and the horizon radius $r_+$ are then all
time-independent, and so is the island contribution \eqref{eq:Sgen_min_classical}. The
only time dependence resides in the no-island branch, through the growth of the conformal
distance between $b_+$ and $b_-$. Once the island saddle takes over, the entropy is
frozen. The same flat plateau appears in every eternal-black-hole island computation,
beginning with Refs.~\cite{Almheiri:2019yqk,hashimoto2020b}, and encodes the statement
that an eternal black hole in equilibrium purifies the bath but does not evaporate.

A descending branch requires the black hole to lose mass. Physically this corresponds to
the one-sided problem in which the ingoing flux is switched off: $r_+$ then decreases, and
because the island saddle tracks the horizon, the plateau
\eqref{eq:Sgen_min_classical} or \eqref{eq:Sgen_min_quantum} becomes a slowly decreasing
function of $t_b$. We now make this quantitative in the adiabatic approximation, which is
the natural extension of the present calculation.

\subsection{Evaporation and the descending branch}
\label{sec:evaporating}

Treating the evaporation as quasi-static, the s-wave Hawking flux of the
two-dimensional matter sector gives
\begin{equation}
\frac{dM}{dt}=-\frac{\pi c}{12}\,T^{2},
\qquad
M(r_+)=\frac{(r_+^{2}+\ell_0^{2})^{3/2}}{2r_+^{2}},
\label{eq:evap_law}
\end{equation}
with $T$ given by Eq.~\eqref{eq:temp}. Near extremality $T\propto\sqrt{\varepsilon}$ and
Eq.~\eqref{eq:evap_law} reduces to $\dot{\varepsilon}=-\varepsilon/\tau$ with
\begin{equation}
\tau=\frac{12\cdot3^{5/2}\pi\,\ell_0^{3}}{c},
\label{eq:evap_time}
\end{equation}
so that $\varepsilon(t)\sim e^{-t/\tau}$. The approach to extremality is exponential and
therefore takes infinite time: the black hole never becomes exactly extremal, but relaxes
onto the extremal configuration asymptotically. This is the precise sense in which the
endpoint is a remnant rather than a naked singularity.

\begin{figure}[t]
	\centering
	\includegraphics[width=0.98\textwidth]{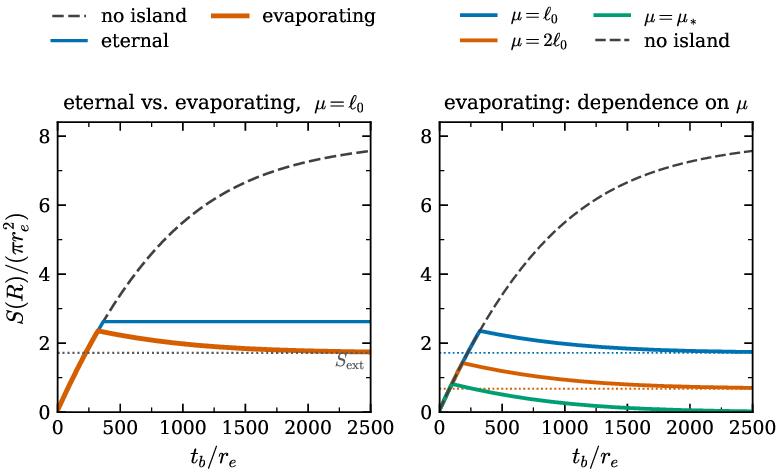}
	\caption{\label{Fig:Page-Curve-evap}Page curve including backreaction, in the
		adiabatic approximation, for an initial $r_0/r_e=0.25$, $b=1.35\,r_e$, $c=100$
		and $\lambda=10^{-2}$, using the corrected entropy functional. Left: the eternal
		configuration (blue) saturates at a constant value, whereas allowing the black
		hole to evaporate (orange) produces a decreasing branch which asymptotes to the
		extremal entropy $S_{\rm ext}$ (dotted), shown here for $\mu=\ell_0$. The dashed
		grey curve is the no-island branch, which also saturates because $\kappa\to0$ as
		extremality is approached. Right: the same evaporating curve for
		$\mu=\ell_0,\,2\ell_0$ and $\mu_\ast=(\sqrt2+\sqrt3)\,\ell_0$, with the
		corresponding asymptotes shown dotted. Changing $\mu$ shifts the island branch
		rigidly and therefore lowers the late-time asymptote, but leaves the no-island
		branch and the shape of the descending branch untouched. At $\mu=\mu_\ast$ the
		asymptote vanishes.}
\end{figure}

Substituting $r_+(t)$ into the island branch, and computing the no-island branch with the
instantaneous surface gravity, $\dot{S}^{\rm (no\;island)}=(c/3)\kappa(t)$, gives the
curves of Fig.~\ref{Fig:Page-Curve-evap}. The Page curve now has the full expected shape:
linear growth, a maximum at $t_{\rm Page}$, and a decreasing branch. The decreasing branch
asymptotes not to zero but to the extremal entropy,
\begin{equation}
\lim_{t\to\infty}S(R)=S_{\rm ext},
\label{eq:late_time_limit}
\end{equation}
because the matter correction in \eqref{eq:Sgen_min_quantum} is suppressed by
$\Lambda_b^{-2}$, which diverges as $r_0\to0$. With the area functional the same limit is
$\pi r_e^{2}$. The residual entropy \eqref{eq:late_time_limit} is exactly the information
that remains locked in the remnant, and the descending branch makes visible what the
eternal calculation could only state indirectly.

Two caveats should be stated. First, the adiabatic treatment is consistent only if the
evaporation is slow compared with the time needed to establish the island saddle. Using
\eqref{eq:evap_time} and \eqref{eq:page_time} one finds that the ratio of the two
timescales is independent of both $c$ and $G$,
\begin{equation}
\frac{\tau}{t_{\rm Page}}\;\simeq\;9\,\frac{r_0}{\ell_0},
\label{eq:adiabaticity}
\end{equation}
so adiabaticity requires $r_0\gtrsim\ell_0/9$. Together with the near-extremality
condition $r_0\ll r_e=\sqrt{2}\ell_0$ this leaves a window
$0.1\lesssim r_0/r_e\lesssim0.5$, which is what we have used in
Fig.~\ref{Fig:Page-Curve-evap}. Outside it --- in particular for the very near-extremal
parameters of Fig.~\ref{Fig:Page-Curve} --- the black hole would evaporate before the
island saddle becomes dominant, and only the eternal configuration, in which the
evaporation is balanced by an ingoing flux, is self-consistent. Second, we have not
included the backreaction of the Hawking flux on the near-horizon geometry itself, which
in the $AdS_2$ throat is governed by the Schwarzian mode and is known to be important at
temperatures below the mass gap~\cite{banerjee2021,iliesiu2025a}.

\subsection{Fixing the renormalization scale}
\label{sec:mu}

The scale $\mu$ in Eq.~\eqref{eq:entropy} is an integration constant: the first law
determines $dS/dr_+$ and hence fixes $S$ only up to an additive constant. Its role is
sharply limited. Since
\begin{equation}
\frac{\partial S_{\rm BH}}{\partial\mu}=-\frac{3\pi\ell_0^{2}}{\mu},
\qquad
\frac{\partial^{2}S_{\rm BH}}{\partial r_+\,\partial\mu}=0,
\label{eq:mu_shift}
\end{equation}
a change of $\mu$ shifts $S_{\rm BH}$ by a constant that is the same for every $r_+$.
Consequently $\mu$ affects neither the extremization condition
\eqref{eq:extremality_quantum}, nor the island location, nor the factor $8/27$ of
Eq.~\eqref{eq:ratio_8_27}, nor the rate at which the descending branch falls. It shifts
the island branch of the Page curve rigidly, and hence shifts the plateau height and the
Page time, as Fig.~\ref{Fig:Page-Curve-evap}(b) shows.

There is nevertheless a sharp constraint on $\mu$. The derivative
\begin{equation}
\frac{dS_{\rm BH}}{dr_+}=2\pi\,\frac{(r_+^{2}+\ell_0^{2})^{3/2}}{r_+^{2}}>0
\label{eq:monotone}
\end{equation}
is strictly positive, so $S_{\rm BH}$ increases monotonically along the physical branch
$r_+\ge r_e$ and attains its minimum at extremality. Demanding $S_{\rm BH}\ge0$ for every
$r_+$ is therefore equivalent to the single condition $S_{\rm ext}\ge0$, which by
Eq.~\eqref{eq:ext_entropy} reads
\begin{equation}
\mu\;\le\;\mu_\ast\;\equiv\;\left(\sqrt{2}+\sqrt{3}\right)\ell_0\;\simeq\;3.146\,\ell_0 .
\label{eq:mu_bound}
\end{equation}
For $\mu>\mu_\ast$ the extremal remnant would carry negative entropy, which is
inadmissible for a von Neumann entropy, and the semiclassical expression
\eqref{eq:entropy} could not be the whole answer.

The boundary of the allowed range is distinguished. At $\mu=\mu_\ast$ one has
$S_{\rm ext}=0$ exactly, so the extremal remnant is a unique quantum state and the Nernst
form of the third law is satisfied: the entropy vanishes as the temperature does. It is
worth emphasising how naturally this arises in the present model. For extremal
Reissner--Nordstr\"om the corresponding statement is obstructed by the residual area term
$A_{\rm ext}/4G$, and reconciling it with the expected absence of a ground-state
degeneracy requires the Schwarzian analysis of
Refs.~\cite{banerjee2021,iliesiu2025a}. Here the area term vanishes on its own, by
Eq.~\eqref{eq:entropy}, and the only obstruction to a zero-entropy ground state is the
choice of a single constant. Fixing $\mu=\mu_\ast$ removes it.

We do not adopt this choice as our default, for the following reason. If
$S_{\rm ext}=0$, the extremal remnant has no accessible states and can store no
information; the entire information content of the initial state must then be returned to
the radiation during the approach to extremality, and the two-stage picture of
Sec.~\ref{sec:discussion} degenerates into the standard one-stage story with an
arbitrarily long tail. Both alternatives are internally consistent, and the choice between
them is a statement about the microscopic theory rather than about the semiclassical
calculation:
\begin{itemize}
\item $\mu<\mu_\ast$: the remnant carries a finite entropy $S_{\rm ext}>0$, information is
	retained in Planckian degrees of freedom, and the Page curve saturates above zero.
\item $\mu=\mu_\ast$: the remnant is a unique state, the third law holds in its strong
	form, and the radiation is asymptotically pure.
\end{itemize}
Discriminating between them requires input the first law cannot supply --- a
microcanonical count of remnant states, or a Euclidean evaluation of the partition
function of the extremal throat. We regard this as the sharpest open question raised by
the present analysis. What the island computation does establish, independently of $\mu$,
is that the plateau of the Page curve is the \emph{full} entropy $S_{\rm BH}(r_+)$ and not
the horizon area, so that whichever value of $\mu$ the microscopic theory selects, it
fixes the information capacity of the remnant unambiguously.

%%%%%%%%%%%%%%%%%%%%%%%%%%%%%%%%%%%%%%%%%%%%%%%%%%%%%%%%%%%%%
\section{Discussion: implications for the information paradox}
\label{sec:discussion}

The results of the previous sections suggest a two-stage picture of information recovery,
organised by the minimal length scale $\ell_0$.

\paragraph{Semiclassical stage, $r_+\gg\sqrt{2}\,\ell_0$.}
Far from extremality the black hole is large and hot, the logarithmic term in
Eq.~\eqref{eq:entropy} is a small correction to the area law, and the island mechanism
operates exactly as in the singular case. The radiation entropy follows a Page curve that
saturates near the Bekenstein--Hawking value, and information is returned to the radiation
in the standard way. Because the temperature \eqref{eq:temp} decreases as $r_+$ approaches
$r_e$, this stage is self-terminating: evaporation slows down as extremality is
approached, and the final approach takes parametrically longer than the earlier evolution.

\paragraph{Quantum stage, $r_+\to\sqrt{2}\,\ell_0$.}
At extremality the temperature vanishes and, as emphasised in Sec.~\ref{sec:geometry},
the area-law piece of the entropy vanishes with it: the prefactor $r_+^{2}-2\ell_0^{2}$ in
Eq.~\eqref{eq:entropy} is precisely the extremality condition. What remains is the purely
logarithmic $S_{\rm ext}$ of Eq.~\eqref{eq:ext_entropy}. The endpoint of evaporation is
therefore not a naked singularity but a cold remnant whose entire entropy --- and hence
whose entire capacity to store information --- resides in quantum-gravitational degrees of
freedom associated with the minimal length. Retrieving that information would require
resolving Planckian modes, on a timescale of order $e^{S_{\rm ext}}$.

It is at this point that the choice of island functional becomes physically significant.
With the standard area term, the Page plateau tends to $\pi r_e^{2}$ in the extremal limit,
Eq.~\eqref{eq:plateau_classical}: the island computation reports a finite area entropy for
a configuration whose thermodynamic entropy contains no area contribution at all. The two
statements are in direct conflict, and the conflict is not small --- for $\mu\sim\ell_0$
the two candidate plateaus differ by a factor of order two. With the corrected functional
the tension disappears: by Eq.~\eqref{eq:plateau_extremal} the plateau is exactly
$S_{\rm ext}$, and the coarse-grained entropy accessible to the island computation
coincides with the entropy obtained from the first law. We regard this consistency as the
main argument in favour of building the island functional from $S_{\rm BH}(a)$ whenever the
underlying theory supplies a corrected entropy, and as an indication that the same question
deserves attention in other regular and higher-curvature black hole models.

Three further consequences are worth recording.

First, the semiclassical island formalism is at its weakest precisely where it matters
most. Both prescriptions rely on a geometric entropy assigned to the island boundary, and
near extremality the classical piece of that entropy vanishes, leaving a term that is
itself a quantum correction. A complete treatment of the final stage would require the
Euclidean path integral including the zero modes of the near-extremal
throat~\cite{banerjee2021,iliesiu2025a}, in which the $AdS_2\times S^2$ geometry of
Eq.~\eqref{eq:ext_ads2_metric} produces its own $\log T$ contributions. Our results should
be read as the semiclassical skeleton of that computation.

Second, the remnant stores information in a form invisible to a low-energy observer. This
offers a resolution of the paradox that does not require the information to be imprinted on
the late-time Hawking quanta, and therefore sidesteps the tension between smoothness of the
horizon and purity of the radiation that motivates firewall arguments.

Third, the logarithmic term is universal in the sense that its coefficient is fixed by
$\ell_0$ alone and is insensitive to most details of the ultraviolet completion. Its
overall normalisation is not: the scale $\mu$ in Eq.~\eqref{eq:entropy} is a renormalisation
constant, and the height of the extremal plateau depends on it. As stressed in
Sec.~\ref{sec:island_quantum}, however, $\mu$ drops out of the extremization entirely, so
the location of the island, the factor $8/27$ in Eq.~\eqref{eq:ratio_8_27}, and the
qualitative shape of the Page curve are all unambiguous. As shown in Sec.~\ref{sec:mu}, positivity of the entropy
restricts $\mu\le\mu_\ast=(\sqrt2+\sqrt3)\ell_0$, the boundary value corresponding to a
remnant of vanishing entropy. Fixing $\mu$ within this range --- ideally by a
microcanonical count of remnant states --- is the natural next step, and would turn
Eq.~\eqref{eq:plateau_extremal} into a prediction for the information capacity of the
extremal remnant.

\section{Conclusion}
\label{sec:conclusion}

We have analysed the black hole information paradox for the T-duality-inspired regular
black hole with minimal length $\ell_0$. Integrating the first law at fixed $\ell_0$ gives
an entropy, Eq.~\eqref{eq:entropy}, consisting of an area-law piece proportional to
$r_+^{2}-2\ell_0^{2}$ and a logarithmic piece proportional to $\ell_0^{2}$. Since
$r_e=\sqrt{2}\ell_0$, the first piece vanishes identically at extremality and the extremal
entropy $S_{\rm ext}$ is purely logarithmic. The extremal near-horizon geometry is
$AdS_2\times S^2$ with radii $\bar{\ell}_0=\sqrt{3}\ell_0$ and $r_e=\sqrt{2}\ell_0$, and
the near-extremal temperature scales as $\sqrt{\varepsilon}$ in the mass above extremality.

Using Kruskal-like coordinates adapted to the near-extremal throat we computed the
radiation entropy without an island, Eq.~\eqref{eq:noisland_final}, which grows linearly
with the surface gravity $\kappa=r_0/\bar{\ell}_0^{2}$ and diverges logarithmically as
extremality is approached. We then computed the island configuration in two ways. With the
standard area functional the island sits a distance
$\sim\lambda^{2}b^{4}/(r_e^{2}r_0\Lambda_b^{6})$ outside the outer horizon --- pinned to
the horizon, as in the Schwarzschild case --- and the Page curve saturates at the extremal
horizon area.

With the entropy functional built from $S_{\rm BH}(a)$ the analysis simplifies rather than
complicating: the variation of the corrected entropy is the algebraic function
$2\pi\Phi(a)$ of Eq.~\eqref{eq:Phi_def}, which is stationary exactly at $r_e$, so that in
the whole near-extremal window the corrected functional reduces to the classical one with
the rescaled coupling $\lambda\to(2/3)^{3/2}\lambda$. The island moves closer to the
horizon by the universal factor $8/27$, the Page time is delayed by
Eq.~\eqref{eq:delta_page_time}, and --- most importantly --- the extremal plateau becomes
exactly $S_{\rm ext}$ rather than $\pi r_e^{2}$. The island computation is then consistent
with the thermodynamics: at extremality all of the entropy, and all of the information, is
carried by the logarithmic term.

Allowing the black hole to evaporate, in the adiabatic approximation, converts the
plateau into a decreasing branch which asymptotes to $S_{\rm ext}$ rather than to zero,
making the remnant interpretation explicit; the approach to extremality is exponential in
time and hence never completed. Positivity of the entropy further bounds the
renormalization scale by $\mu\le(\sqrt{2}+\sqrt{3})\ell_0$, the boundary value being the
unique choice for which the extremal remnant is a zero-entropy ground state.

Several directions suggest themselves. The substitution $r_+\to a$ used here is justified
only if the logarithmic term descends from a local covariant entropy functional; deriving
such a functional directly from the T-duality-inspired action of Ref.~\cite{nicolini2019},
or showing that it does not exist, would settle which prescription is correct. Fixing the
renormalisation scale $\mu$ by a microscopic count would convert the extremal plateau into
a quantitative statement about the remnant. Finally, it would be interesting to repeat the
analysis for other regular black holes --- Bardeen and Hayward in particular --- to
determine whether the vanishing of the area term at extremality, which is what makes the
choice of functional matter here, is a generic feature of minimal-length geometries or an
accident of this model.

\section*{Acknowledgements}
Ankit Anand is financially supported by the Institute's postdoctoral fellowship at IIT Kanpur. Amir A.~Khodahami and Ahmad Sheykhi would like to thank Shiraz University Research Council. The authors acknowledge the use of artificial intelligence tools for assistance with improving the presentation of the manuscript.

%----------------------------
%\bibliographystyle{utphys}
%\begin{thebibliography}{99}
%\bibliographystyle{apsrev4-2}
\bibliography{18th_paper}
%\end{thebibliography}

\end{document}